\documentclass[10pt,superscriptaddress,amsmath,amssymb,aps,prl,twocolumn,floatfix,longbibliography]{revtex4-1}

\usepackage[version=3]{mhchem}
\usepackage{graphicx,times,bm,amssymb,amsmath,booktabs,natbib,physics,subfigure,makecell,hyperref,multirow}
\usepackage{braket}
\usepackage{microtype}

\usepackage[version=3]{mhchem}
\usepackage{physics,subfigure,makecell,hyperref,multirow,color}
\usepackage{times,color}
\usepackage{amsfonts}
\usepackage{mathrsfs}

\hypersetup{colorlinks,allcolors=black}

\def\be{\begin{equation}}       \def\ee{\end{equation}}
\def\bea{\begin{eqnarray}}      \def\eea{\end{eqnarray}}

\usepackage{graphicx}
\usepackage{epstopdf}

\begin{document}
\title{Ultrafast optical route to coupled ferroelectric and altermagnetic switching}

\author{Yuhao Gu}
\affiliation{School of Mathematics and Physics, University of Science and Technology Beijing, Beijing 100083, China}

\author{Yu-hui Song}
\affiliation{ School of Physics and Beijing Key Laboratory of Opto-electronic Functional Materials $\&$ Micro-nano Devices. Renmin University of China, Beijing 100872, China}
\affiliation{ Key Laboratory of Quantum State Construction and Manipulation (Ministry of Education), Renmin University of China, Beijing 100872, China}

\author{Peng-Jie Guo}  \email{guopengjie@ruc.edu.cn}
\affiliation{ School of Physics and Beijing Key Laboratory of Opto-electronic Functional Materials $\&$ Micro-nano Devices. Renmin University of China, Beijing 100872, China}
\affiliation{ Key Laboratory of Quantum State Construction and Manipulation (Ministry of Education), Renmin University of China, Beijing 100872, China}

\author{Yihao Wang}
\affiliation{School of Mathematics and Physics, University of Science and Technology Beijing, Beijing 100083, China}

\author{Zhe Li}
\affiliation{School of Mathematics and Physics, University of Science and Technology Beijing, Beijing 100083, China}

\author{Ze-Feng Gao}
\affiliation{ School of Physics and Beijing Key Laboratory of Opto-electronic Functional Materials $\&$ Micro-nano Devices. Renmin University of China, Beijing 100872, China}
\affiliation{ Key Laboratory of Quantum State Construction and Manipulation (Ministry of Education), Renmin University of China, Beijing 100872, China}

\author{Huan-Cheng Yang}
\affiliation{ School of Physics and Beijing Key Laboratory of Opto-electronic Functional Materials $\&$ Micro-nano Devices. Renmin University of China, Beijing 100872, China}
\affiliation{ Key Laboratory of Quantum State Construction and Manipulation (Ministry of Education), Renmin University of China, Beijing 100872, China}

\author{Zhong-Yi Lu} \email{zlu@ruc.edu.cn}
\affiliation{ School of Physics and Beijing Key Laboratory of Opto-electronic Functional Materials $\&$ Micro-nano Devices. Renmin University of China, Beijing 100872, China}
\affiliation{ Key Laboratory of Quantum State Construction and Manipulation (Ministry of Education), Renmin University of China, Beijing 100872, China}
\affiliation{ Hefei National Laboratory, Hefei 230088, China}

\begin{abstract}

Exploring novel magnetoelectric coupling mechanisms to achieve control of ferroelectric polarization and magnetism is highly significant for both fundamental science and electronic device applications. Although extensive studies have been conducted on electrical switching of magnetism in multiferroic materials, simultaneous ultrafast laser switching of ferroelectric polarization and altermagnetism remains unexplored. In this letter, we propose that the ultrafast laser can be used to switch ferroelectric polarization and altermagnetism concurrently in charge-order-induced altermagnetic ferroelectrics. Building on this idea, we further demonstrate that such dual switching can be realized in charge-order-induced altermagnetic ferroelectric LiV$_2$F$_6$ by symmetry analysis and time-dependent density functional theory (TDDFT) calculation. Given that LiV$_2$F$_6$ has already been experimentally synthesized, our work not only provides an ideal material platform for experimentally realizing simultaneous switching of ferroelectric polarization and altermagnetism but also holds potential application value in future ultrafast spintronic devices.

\end{abstract}

\maketitle

Altermagnets combine compensated magnetic order with momentum-dependent spin splitting \cite{li-prx, JPS, ha-prb, liber-prx}. This unusual combination has motivated proposals for spin-split currents \cite{Liber-PRL,Bai-PRL, Ka-PRL}, piezomagnetism \cite{piezomagnetism-NC}, anomalous transport \cite{CAHE-2020, MnTe-PRL, Liber-NRM, Feng-NE, hou-prb}, and topological responses \cite{PhysRevB-tan,BWS,feng2025typeiiquantumspinhall,tan2025stackinginducedtypeiiquantumspin}. In practical applications, a central challenge is how to switch the spin-split electronic structure without introducing a net magnetization.

Ferroelectric altermagnets offer a route for such control: reversing the ferroelectric polarization can concurrently reverse the altermagnetic spin splitting while preserving compensated magnetism \cite{guo2023altermagnetic, Ferroelectric-1, Ferroelectric-2,vsmejkal2024altermagnetic,Sun2025AdvMater,nju_guo2025altermagnetic,rabe2025BFOalter}. The microscopic origin of this coupling is crucial. If ferroelectricity and magnetism arise from unrelated order parameters, the magnetoelectric response is generally weak. A stronger route is to make both orders arise from the same electronic instability.

\begin{figure}[b]  \centerline{\includegraphics[width=0.5\textwidth]{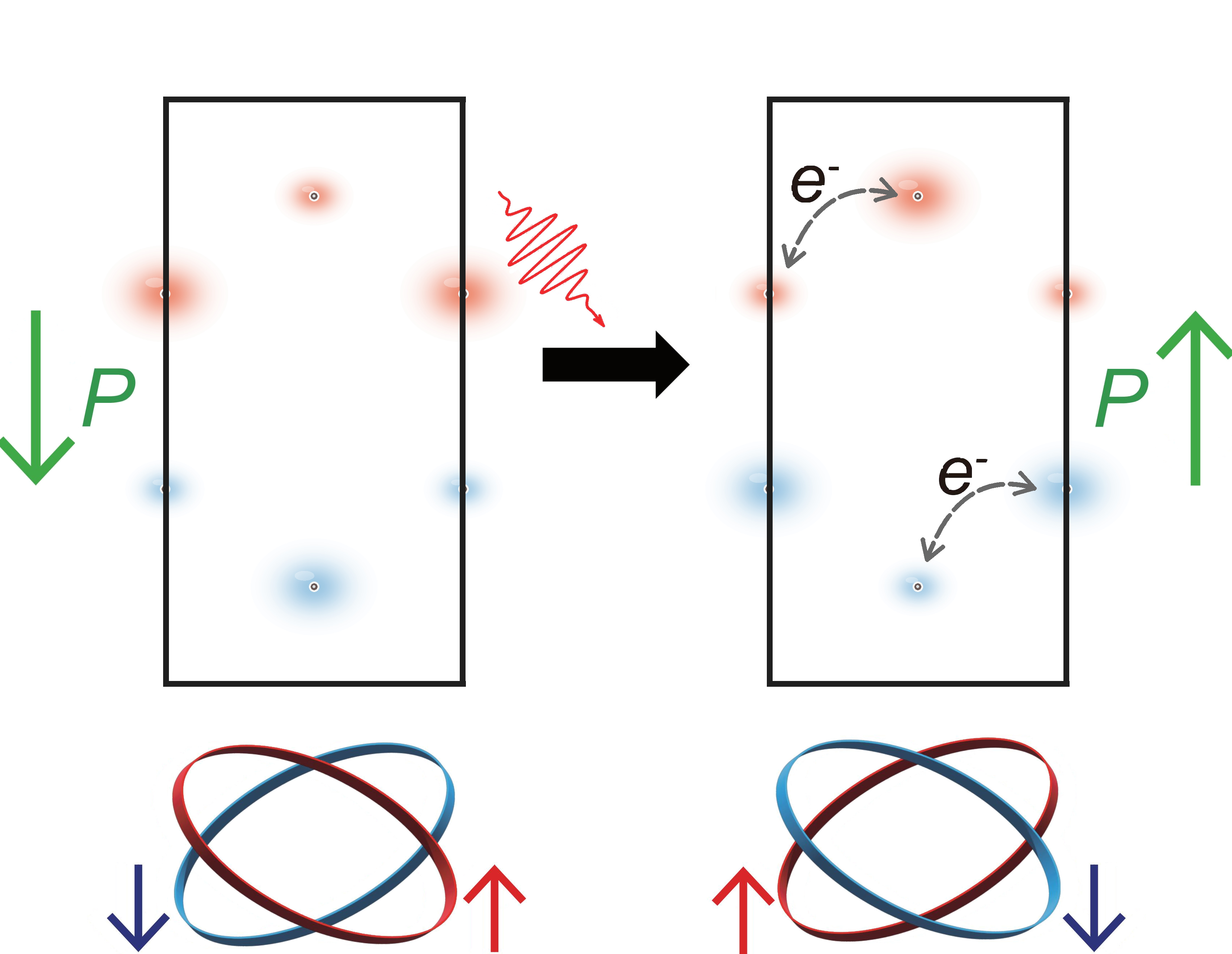}}
\caption{Schematic illustration of manipulating altermagnetism and reversing ferroelectric polarization simultaneously by laser or external electric field in charge-order-induced ferroelectrics. The green arrow represents the direction of ferroelectric polarization. The red and blue arrows represent spin up and down, respectively. The Fermi surfaces and electron clouds are also marked by red and blue for spin up and down.  
		\label{fig1} }
\end{figure}

Ferroelectricity can originate from several microscopic mechanisms, such as displacive lattice distortions and lone-pair electrons \cite{ishihara2010electronic,yamauchi2014electronic,khomskii2006multiferroics,hill2000why}. Charge-order-induced ferroelectricity is particularly attractive because its polarization reversal can, in principle, be driven by electron hopping rather than by lattice-mediated ionic motion, potentially enabling much faster switching than that in conventional ferroelectrics \cite{qi2022electron,park2017charge,park2019superlattice}. However, as charge carriers easily fluctuate around atomic sites, the practical implementation of such materials is often hindered by leakage currents and dielectric breakdown \cite{qi2022electron,alexe2009ferroelectric}. Moreover, strong coupling between charge order and lattice relaxation could lock a particular charge-ordering state, making purely electrical switching energetically unfavorable and in some cases requiring unrealistically large fields beyond the dielectric breakdown limit \cite{qi2022electron}.

These features make optical control especially appealing for charge-order-induced multiferroics. Because polarization reversal is fundamentally governed by charge transfer rather than by ionic displacement, light can directly trigger the relevant charge redistribution on ultrafast timescales in a contactless manner \cite{subedi2015proposal,mankowsky2017ultrafast,chen2022deterministic,yang2024light,wang2024ultrafast}. On the other hand, in conventional multiferroics, where polarization reversal is mainly controlled by atomic motion, simultaneous optical control of ferroelectric polarization and magnetism is generally more difficult. This approach is particularly important for strongly coupled altermagnetic multiferroics, in which charge order can provide a common microscopic origin for both ferroelectricity and altermagnetism. Naturally, optical excitation therefore offers a contactless and ultrafast route to the coupled switching of ferroelectric polarization and altermagnetic spin polarization in such systems.

In this Letter, we propose such a mechanism: charge order simultaneously induces ferroelectricity and altermagnetism. Since both ferroelectricity and altermagnetism originate from charge order, such systems exhibit strong magnetoelectric coupling and enable electrical control of altermagnetism. Specifically, ferroelectricity arises from different valence states of the same magnetic atoms, so that electron hopping between magnetic atoms can induce ferroelectric polarization reversal and consequently switch altermagnetism (as illustrated in Fig.~\ref{fig1}). Such reversal may in principle be driven electrically, but ultrafast laser excitation provides a contactless route to trigger the underlying charge redistribution and reduce issues associated with leakage currents and dielectric breakdown. Based on this scenario, through symmetry analysis and first-principles calculations, we predict LiV$_2$F$_6$ to be a charge-order-induced altermagnetic ferroelectric material. Furthermore, using TDDFT, we demonstrate that ultrafast laser pulses can simultaneously switch ferroelectric polarization and altermagnetism. The computational methods are detailed in our Supplemental Material (SM); see also references \cite{kresse1996,Joubert1999,perdew_generalized_1996,liechtenstein1995,metzger1983crystal,baroni2001,togo2015,togo2023,king1993,resta1994,NEB,tancogne2020octopus,dudarev1998electron} therein.

\begin{figure}[t]	\centerline{\includegraphics[width=0.5\textwidth]{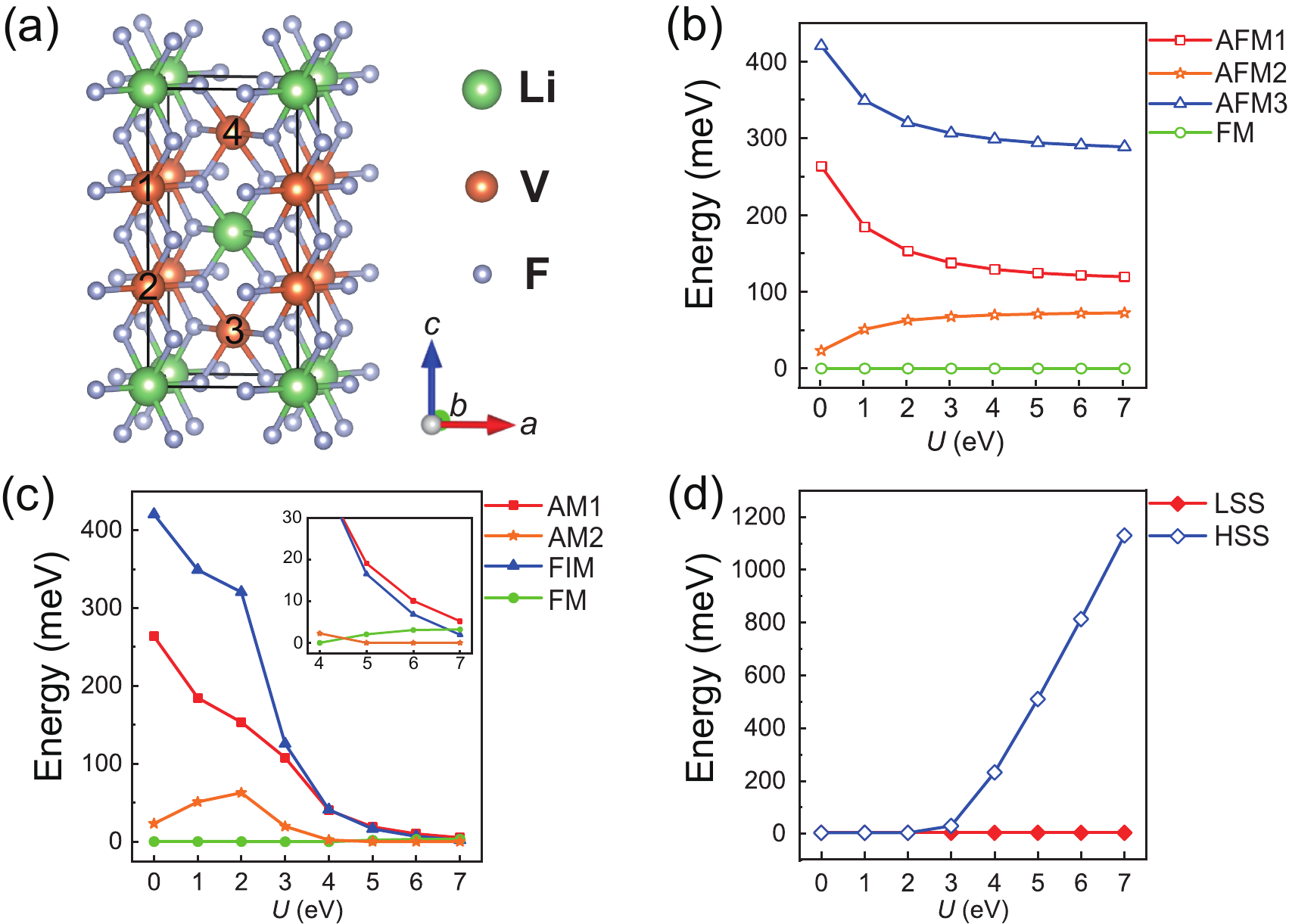}} 
 \caption{(a) Crystal structure of HSS \ce{LiV2F6} at room temperature. The green/orange/blue-gray ball represents Li/V/F atom, respectively. (b) Energies for different magnetic orders as a function of $U$ in HSS \ce{LiV2F6}. (c) Energies for different magnetic orders as a function of $U$ in LSS \ce{LiV2F6}. (d) The comparison of ground state energies of HSS and LSS structure as a function of $U$. The ground state energy is set as 0 with every $U$ value in (b/c/d).
		\label{fig2} }
\end{figure}

%


Based on the principle for coupled charge-order, ferroelectric, and altermagnetic switching shown in Fig.~\ref{fig1}, the magnetic atoms in the corresponding high-symmetry structure should have fractional valence states. This condition favors the emergence of charge order as temperature decreases, subsequently leading to ferroelectricity. 

We propose that LiV$_2$F$_6$ could be an appropriate candidate material: the average valence state of V atoms is +2.5, and the reported temperature-dependent magnetic susceptibility of \ce{LiV2F6} exhibits an antiferromagnetic character \cite{metzger1983crystal}. The experimentally reported room-temperature structure of \ce{LiV2F6} \cite{metzger1983crystal}, illustrated in Fig.~\ref{fig2}(a), is a typical trirutile lattice where Li and V alternatively occupy 1/3 and 2/3 of the MF$_6$ octahedra along the $z$ axis. In this structure, all V atoms are equivalent and the space group is $P4_2/mnm$, with the corresponding point group D$_{4h}$, which includes generators C$_{4z}$(1/2, 1/2, 1/2), C$_{2x}$(1/2, 1/2, 1/2), and $I$. Since LiV$_2$F$_6$ has a nonsymmorphic space group, its unit cell contains four V atoms (Fig.~\ref{fig2}(a)). Below we refer to this experimentally reported paraelectric structure as the high-symmetry structure (HSS).

To determine the magnetic ground state of LiV$_2$F$_6$, we consider four typical magnetic structures, including three antiferromagnetic (AFM1, AFM2, AFM3) and one ferromagnetic (FM) configurations, as detailed in our SM. Clearly, AFM1 possesses $\{E\|I\}$ spin symmetry but lacks $\{C_{2}^{\bot}\|I\}$ spin symmetry. Due to the presence of F atoms, the V atoms with opposite spin magnetic moments cannot be connected by the fractional translation of (1/2, 1/2, 1/2), and thus the AFM1 state also lacks $\{C_{2}^{\bot}\|(1/2, 1/2, 1/2)\}$ spin symmetry. Considering that the V atoms with opposite spin magnetic moments can be related by C$_{4z}$(1/2, 1/2, 1/2) symmetry, the AFM1 is also a $d$-wave altermagnetic state,  labeled as AM1 in Fig.~S1(a). In both AFM2 and AFM3, the V atoms with opposite spin magnetic moments can be connected by inversion symmetry, and therefore both are conventional antiferromagnets (Figs.~S1(b-c)). Then, using the GGA plus on-site repulsion $U$ method (GGA+$U$), we calculate the relative energies of these four magnetic structures as a function of the on-site Coulomb interaction $U$, as shown in Fig.~\ref{fig2}(b). From Fig.~\ref{fig2}(b), the FM state remains the most stable across different values of $U$, which is inconsistent with previous experimental results \cite{metzger1983crystal}.

\begin{figure}[t]
	 \centerline{\includegraphics[width=0.5\textwidth]{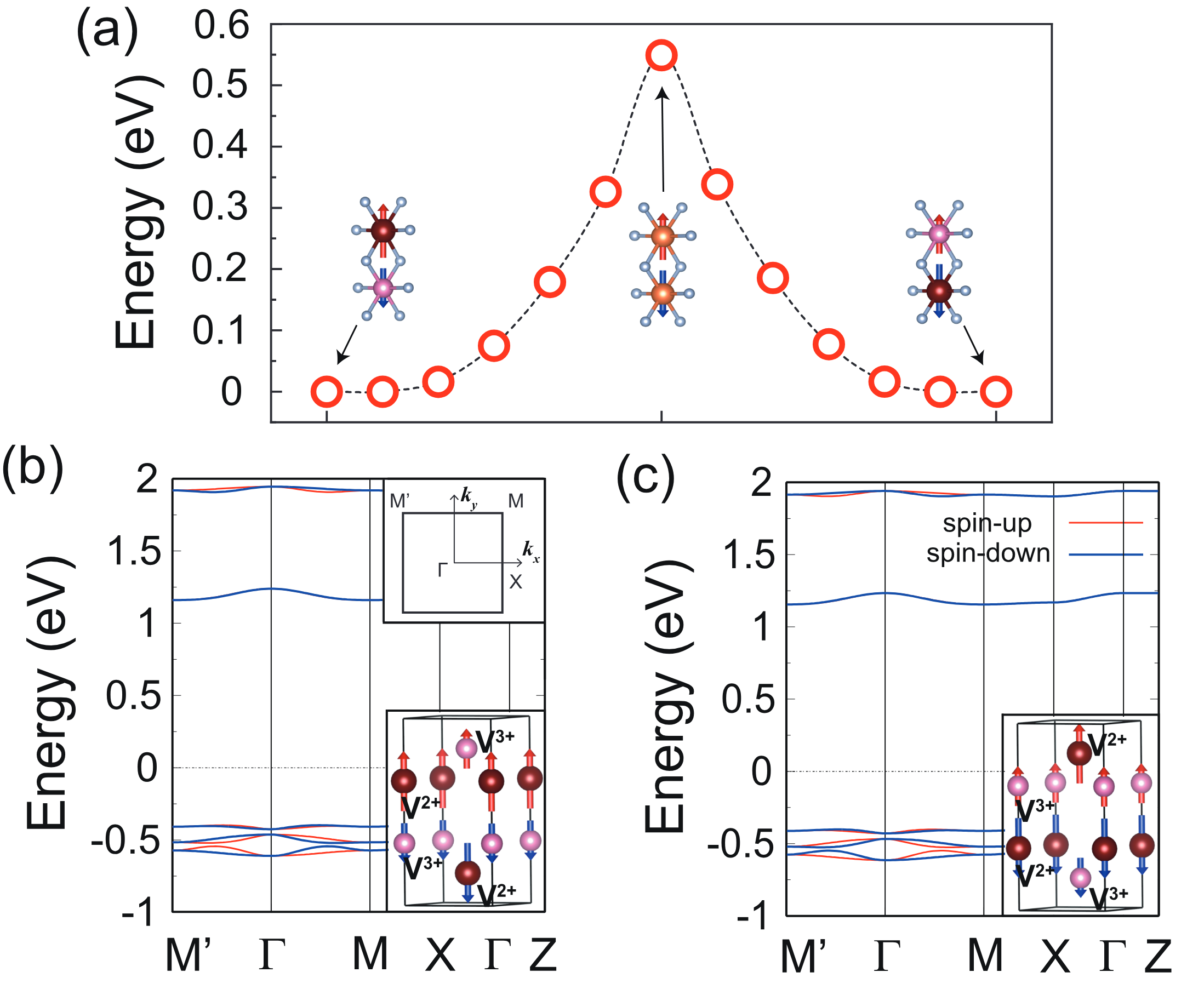}}
 \caption{(a) The NEB-simulated energies in the process of switching ferroelectric polarization. The insets sketch initial, intermediate and final structures. The dark red/orange/pink ball represents \ce{V^{2+}}/\ce{V^{2.5+}}/\ce{V^{3+}}, respectively.
 (b-c) Electronic band structures of initial and final structures. The top-right inset in (b) illustrates the first Brillouin zone and the high-symmetry $\textbf{k}$-points. The other two bottom-right insets illustrate the charge and magnetic order of the initial and final structures, respectively. 
 \label{fig3} }
\end{figure}

Many mixed-valence vanadium compounds exhibit charge ordering or valence disproportionation \cite{nakao2000xray,park2017charge,yamauchi2005charge}. Motivated by the mixed V valence and by the discrepancy between the calculated HSS ferromagnetic ground state and the reported antiferromagnetic susceptibility, we consider a possible charge-ordered phase of LiV$_2$F$_6$ as a theoretical prediction to be tested experimentally. We therefore investigate the magnetic properties of LiV$_2$F$_6$ in a broken-inversion-symmetry structure induced by charge order, which we denote below as the low-symmetry structure (LSS). 
Our calculations on the four magnetic structures of LSS LiV$_2$F$_6$ show the emergence of V$^{2+}$ and V$^{3+}$, that is, the charge order, as shown in Fig.~S2. For AFM1 (AM1), although the emergence of V$^{2+}$ and V$^{3+}$ breaks the inversion symmetry, it still possesses $\{C_{2}^{\bot}\|C_{4z}(1/2, 1/2, 1/2)\}$  spin symmetry that connects atoms with opposite spin magnetic moments. Therefore, it remains to be a $d$-wave altermagnetic state (Fig.~S2(a)). Unlike AFM1, the appearance of V$^{2+}$ and V$^{3+}$ breaks $\{C_{2}^{\bot}\|I\}$ spin symmetry, and the magnetic atoms with opposite spin magnetic moments can be connected by C$_{4z}$(1/2, 1/2, 1/2) symmetry, causing AFM2 to transform into the $d$-wave altermagnetic state AM2 (Fig.~S2(b)). Evidently, due to the emergence of V$^{2+}$ and V$^{3+}$, the total magnetic moment in AFM3 becomes non-zero, thus transforming it into a ferrimagnetic (FIM) state (Fig.~S2(c)).

We next compare the energetics and dynamical stability of the two structures. The charge-ordered LSS is lower in energy for $U$ values above 2 eV (Fig.~\ref{fig2}(d)). Moreover, when $U\ge 5$ eV, the magnetic ground state changes from the HSS ferromagnetic state to the LSS altermagnetic AM2 state, consistent with the reported antiferromagnetic susceptibility \cite{metzger1983crystal}. We therefore use $U=5$ eV in the following calculations as the value constrained by comparison with the experimental magnetic character. Phonon calculations provide the dynamical stability check: the paraelectric HSS exhibits unstable phonon modes, whereas the charge-ordered ferroelectric LSS has no imaginary phonon frequency (Fig. S3). These results identify the charge-ordered LSS as a predicted stable low-symmetry phase in which space inversion breaking induces altermagnetism.

In LSS LiV$_2$F$_6$, the alternating V$^{2+}$ and V$^{3+}$ ions induce ferroelectricity, similar to the scenario in \ce{LiFe2F6} \cite{lin2017ferroelectric,dong2019magnetoelectricity,guo2023altermagnetic}. The Berry-phase polarization is $P=12.1$ $\mu C/cm^2$ along the $z$ axis for the altermagnetic AM2 state, consistent with a classical point-charge estimate of 10.6 $\mu C/cm^2$. The climbing image nudged elastic band (CI-NEB) calculation \cite{NEB} gives a polarization-reversal barrier of about 0.55 eV/f.u. [Fig.~\ref{fig3}(a)], larger than that of \ce{LiFe2F6} \cite{lin2017ferroelectric,guo2023altermagnetic} and comparable to that of \ce{BiFeO3} \cite{ravindran2006BiFeO3}. In this NEB calculation, the magnetic structure is fixed as AM2, although FM is the ground state magnetic structure in HSS \ce{LiV2F6}. Therefore, LiV$_2$F$_6$ exhibits charge-order-induced ferroelectricity. 

More importantly, the charge-order-induced ferroelectricity also causes LiV$_2$F$_6$ to transition from the FM state to the altermagnetic AM2 state, indicating that LiV$_2$F$_6$ exhibits strong magnetoelectric coupling. Therefore, polarization-controlled altermagnetic spin splitting may be achievable in LiV$_2$F$_6$. To demonstrate this coupling, we first calculate the electronic band structure of LiV$_2$F$_6$ with upward ferroelectric polarization, as shown in Fig.~\ref{fig3}(b). From Fig.~\ref{fig3}(b), it is evident that LiV$_2$F$_6$ is an altermagnetic semiconductor. Due to the absence of spin symmetry $\{C_{2}^{\bot}\|I\}$ and $\{C_{2}^{\bot}\|(1/2, 1/2, 1/2)\}$, LiV$_2$F$_6$ has $k$-dependent spin splitting. In fact, due to the spin symmetry $\{C_{2}^{\bot}\|M_{x}(1/2, 1/2, 1/2)\}$ and $\{C_{2}^{\bot}\|M_{y}(1/2, 1/2, 1/2)\}$, spin-up and spin-down bands are degenerate on the four planes where k$_x$ and k$_y$ equal 0 or $\pi$ (Fig.~\ref{fig3}(b)). Except for these four faces, spin-up and spin-down bands are split at general $k$ points in the Brillouin zone; for example, along the high-symmetry M'--$\Gamma$--M path (Fig.~\ref{fig3}(b)), this also reflects the characteristic features of $d$-wave altermagnetism. Then, we calculate the band structure of LiV$_2$F$_6$ with downward ferroelectric polarization, as shown in Fig.~\ref{fig3}(c). Comparing Fig.~\ref{fig3}(b) and Fig.~\ref{fig3}(c), the band structure features of LiV$_2$F$_6$ with upward and downward ferroelectric polarization are identical, but the spin polarization of the bands is reversed. This establishes a one-to-one correspondence between polarization reversal and altermagnetic spin-polarization reversal, without reversing a net magnetization.

\begin{figure*}[t]
	 \centerline{\includegraphics[width=0.9\textwidth]{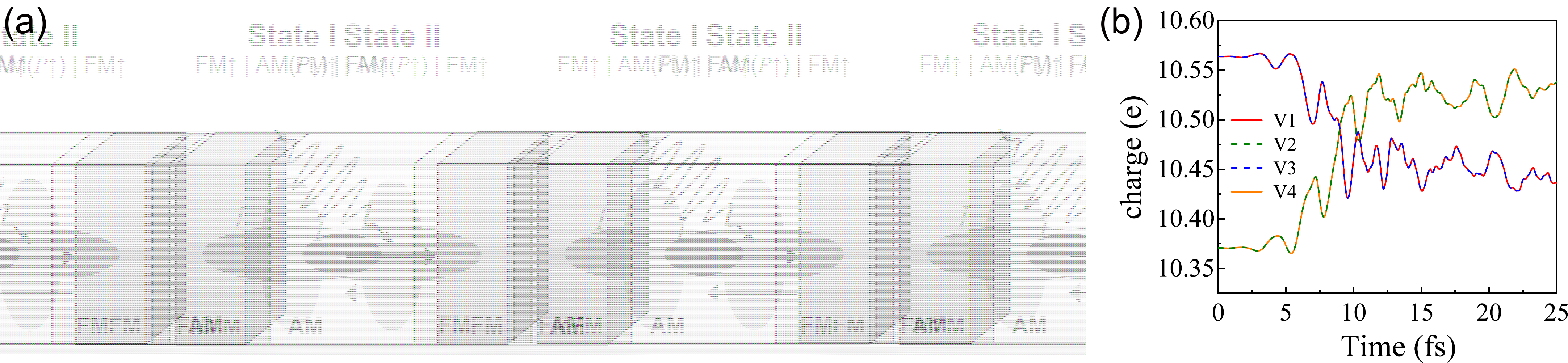}}
 \caption{(a) Schematic illustration of a possible optically assisted altermagnetic ferroelectric spintronic device. Reversing the charge-order-driven polarization of the central layer switches the altermagnetic spin polarization, leading to a change in the spin-dependent conductance while the magnetization directions of the two ferromagnetic electrodes remain unchanged. (b) Time evolution of the site-resolved charges on the four V atoms under laser excitation. The solid lines denote the spin-up channel (V1 and V4), whereas the dashed lines denote the spin-down channel (V3 and V2).
 \label{fig4} }
\end{figure*}

Charge-order-induced ferroelectrics often face problems of leakage currents and dielectric breakdown \cite{alexe2009ferroelectric,qi2022electron}. Thus, light may provide a contactless and ultrafast route to drive charge redistribution in such systems \cite{subedi2015proposal,mankowsky2017ultrafast,chen2022deterministic,yang2024light,wang2024ultrafast}. Inspired by previous proposals for ferroelectric switchable altermagnetic spintronic devices \cite{Ferroelectric-2,zhouTong2025device}, Fig.~\ref{fig4}(a) illustrates a possible device consequence of the coupled switching mechanism. In this geometry, the two ferromagnetic electrodes serve as fixed spin polarizer and analyzer, whereas the central LiV$_2$F$_6$ layer acts as a switchable altermagnetic ferroelectric barrier. Reversal of the charge-order-driven polarization would switch the system between the AM($P{\downarrow}$) and AM($P{\uparrow}$) states, thereby reversing the band spin polarization and changing the spin-matching condition with the electrodes. Consequently, the spin-dependent conductance could be modulated without reversing the magnetization directions of the ferromagnetic electrodes. It is worth noting that the two ferromagnetic electrodes here can also be replaced with two non-polar altermagnetic electrodes.

We then simulate the photoinduced ultrafast charge-transfer dynamics in LiV$_2$F$_6$ under laser excitation with the TDDFT method. Fig.~\ref{fig4}(b) displays the time evolution of charges on the four distinct vanadium sites for both spin channels. For the spin-up channel  (V1 and V4, red and orange solid lines), the charge on the V1 (formal valence $+2$) decays rapidly from 10.56~$e$ to 10.43~$e$ within 12~fs, and then oscillates around 10.43~$e$ in the 12--25~fs time window. Concurrently, the charge on the V4 (formal valence $+3$) shows a complementary increase from 10.37~$e$ to 10.54~$e$. This anti-phase behavior constitutes a direct charge transfer from V1 to V4. Symmetrically, an identical and synchronized charge transfer process is observed for the spin-down channel (V3 and V2, green and blue dashed lines) from V3 to V2. The combined dynamics in both spin channels demonstrate a net displacement of charge between the different valence states, as schematically illustrated in Figs.~\ref{fig3}(b) and \ref{fig3}(c). Because the two static polarization states have opposite altermagnetic spin polarization, this femtosecond charge redistribution identifies the microscopic optical pathway toward coupled ferroelectric and altermagnetic switching in LiV$_2$F$_6$.

The mechanism found in LiV$_2$F$_6$ suggests two design rules for optically assisted coupled switching of ferroelectric polarization and altermagnetic spin polarization. First, the high-symmetry structure should contain magnetic atoms with fractional valence states, favoring charge ordering. Second, the charge-order-induced inversion breaking should create altermagnetism that is absent in the high-symmetry phase. For example, LiFe$_2$F$_6$ exhibits altermagnetism both before and after charge ordering \cite{guo2023altermagnetic}, and therefore does not show the same charge-order-enabled magnetoelectric switching. These two conditions provide a practical guide for searching for altermagnetic multiferroics with optically driven charge-order dynamics.

The predicted charge-ordered ferroelectric phase can be tested experimentally by searching for low-temperature symmetry lowering, inversion-symmetry breaking, and V-valence disproportionation in LiV$_2$F$_6$. The predicted reversal of altermagnetic spin polarization provides a further target for spin-sensitive probes or spin-dependent transport in suitably prepared samples.

In summary, we identify charge order as a microscopic switch that can reverse ferroelectric polarization and altermagnetic spin polarization together. LiV$_2$F$_6$ realizes this principle in first-principles calculations: the charge-ordered polar phase is dynamically stable, altermagnetic, and connected to its opposite-polarization partner by reversed spin splitting. Real-time TDDFT calculation further shows that ultrafast laser excitation drives femtosecond charge transfer along this reversal pathway. These results establish a route to optical manipulation of coupled ferroelectric and altermagnetic orders and provide concrete design principles for altermagnetic multiferroics.

{\it Acknowledgement:}
We thank Shuai Dong, Ning Ding and Yali Yang for useful discussions. The work is supported by the National Science Foundation of China (Grant No.12404153, No.12434009), the National Key R$\&$D Program of China (Grant No. 2024YFA1408601), the Fundamental Research Fund for the Central Universities (Grant No.FRF-TP-25-040) and the Fundamental Research Funds for the Central Universities, and the Research Funds of Renmin University of China (Grant No. 24XNKJ15). Computational resources have been provided by the Physical Laboratory of High Performance Computing at Renmin University of China and Hefei advanced computing center.

Y. G., Y.-H. S. and P.-J. G. contributed equally to this work.

\bibliography{ref}
\end{document}